\input epsf
\documentclass[twocolumn,psfig,aps,prl]{revtex4}
\begin{document}
\pagenumbering{arabic} 

\newcommand{\zero}    {^{(0)}}
\newcommand{\one}     {^{(1)}}
\newcommand{\two}     {^{(2)}}

\title{An efficient  ${\bf k} \cdot {\bf p} $ method for the calculation of total energy and electronic density of states}
\author{Marcella Iannuzzi and Michele Parrinello}  
\affiliation{CSCS - Swiss Center for Scientific Computing, via Cantonale,
 CH-6928 Manno  \\
and \\
Physical Chemistry ETH, H\"onggerberg HCI, CH-8093 Zurich, Switzerland.
}

\begin{abstract}
An efficient method for calculating the electronic structure in large systems   with a fully converged BZ sampling is presented. The method is based on a ${\bf k}\cdot {\bf p}-$like approximation developed in the framework of the density functional perturbation theory. The reliability and efficiency of the method are demostrated in test calculations on Ar and Si supercells.  
\end{abstract}
\maketitle

In periodic systems the  correct description of electronic structure requires an exhaustive sampling of the Brillouin Zone (BZ). However for large systems, such as those that are  employed in {\it ab-initio} molecular dynamics, this can  be too expensive and very often the sampling is limited to the $\Gamma$ point of the BZ.
This is quantitatively accurate for very large systems and/or wide gap insulators.
In all the other cases a correct sampling of the BZ is to be recommended.
Another limitation of $\Gamma$ calculations is the very coarse description that they give of  the electronic density of states, even for rather large supercells.

However, since full BZ sampling can be too costly, ${\bf k} \cdot {\bf p}-$like approaches have been suggested by several authors  in order  to include ${\bf k} \ne  0$ states at reduced computational costs \cite{kdp1}\cite{kdp2}.
In these works it has  been suggested that the calculation of the ${\bf k}$ dispersion of eigenvalues can be conducted in a restricted Hilbert space.
Robertson and Payne \cite{kdp1} have proposed that  this Hilbert space  could be constructed from the occupied and a limited number of unoccupied eigenfunctions of the $\Gamma$ only KS Hamiltonian. 
Since by increasing the number of unoccupied states one converges  to the exact result, this method is in principle exact.
Unfortunately,  the number of excited states that needs to be included in order to achieve  a satisfactory convergence is large, therefore  this  approach has had limited impact.
Kohanoff and Scandolo \cite{kdp2} have proposed a procedure in which the basis vectors of the Hilbert space are variationally improved, along with the electronic wavefunction coefficients. Still, in the non-local PP formalism, this procedure would become rather expensive and actually it has been applied only to the local PP, which is quite restrictive.

Here we adapt a recently developed version of density functional perturbation theory (DFPT) \cite{Baroni}\cite{Gonze}\cite{Putrino} to calculate the finite ${\bf k}$ corrections. We show that this can give very accurate energies for systems of a few hundred atoms for a semiconductor like Si, and even smaller for an insulator like Ar, producing a significant improvement of the $\Gamma $ only sampling at reduced computational cost. 
Furthermore we show that for large systems,  the Hilbert space of the occupied eigenstates supplemented by the perturbative corrections provide an excellent description of the electronic density of states (DOS) at a much reduced computational cost.

We shall restrict ourselves to the case of a semiconductor or insulator, where all states $\psi_{{\bf k}m}({\bf r})$ labeled by the state index $m$ and by the BZ vector ${\bf k}$ are doubly occupied.
 We shall develop our method within the context of pseudopotential (PP) formalism, but it can  easily be extended to other electronic structure schemes. In order to be definite we shall consider fully non-local PP of the Kleinman-Beylander type \cite{pseudo} composed of a local part, $V_{loc}({\bf r})$, and a non-local part, $\sum_{I,L}|P_{I,L}\rangle\omega_{L}\langle P_{I,L}|$, where the sum runs over all nuclei $I$ and the angular momentum channels $L$.
 The Bloch theorem \cite{Ashcroft} allows us to write   $\psi_{{\bf k}m}({\bf r}) = u_{{\bf k}m}({\bf r})e^{i{\bf k}\cdot{\bf r}}$ where  $u_{{\bf k}m}({\bf r})$ has the lattice periodicity.
In terms of the $u_{{\bf k}m}$ the Kohn and Sham (KS) density functional \cite{DFT} can be rewritten as
 \begin{eqnarray}
E_{KS} & = & \sum_{BZ}E({\bf k})
\end{eqnarray}
where the contribution from each ${\bf k}$ vector reads
\begin{eqnarray}
E({\bf k}) & = & 
 2 \sum_{m}^{occ}\Big\{ \langle u_{{\bf k}m}|-\frac{1}{2}\nabla^{2}|u_{{\bf k}m} \rangle + \frac{1}{2}{\bf k}^{2} \nonumber \\
 & & + \langle u_{{\bf k}m}|-i {\bf k}\cdot{\bf \nabla}| u_{{\bf k}m} \rangle  \nonumber \\
  & +  &  \sum_{I,L}\Big[\langle u_{{\bf k}m}\exp{(i{\bf k}\cdot{\bf r})}|P_{I,L}\rangle \omega_{L}  \nonumber \\
  &  &  \; \langle P_{I,L}| u_{{\bf k}m}\exp{(i{\bf k}\cdot{\bf r})}\rangle \Big] \Big\}  \nonumber  \\
  & +  & \int \rho({\bf r})V_{loc}({\bf r})d{\bf r} +E_{Hxc}(\rho({\bf r}))
\label{eq: KSeq}
\end{eqnarray}
The sum in m is over the occupied states.   The electronic charge density is given by 
\begin{equation}
\rho({\bf r}) = 2\sum_{m}\sum_{BZ}|u_{{\bf k}m}({\bf r})|^{2},
\end{equation}
$E_{Hxc}$ is the sum of the Hartree and exchange and correlation contributions to the energy density functional.
In the definition of the  energy functional  the orthonormality condition is implied
\begin{equation}
\langle u_{{\bf k}m} | u_{{\bf k}n}\rangle = \delta_{mn} 
\end{equation}

For large unit cells, the BZ is very small and  the functional in Eq.(\ref{eq: KSeq}) can be expanded in  ${\bf k}$. 
The ${\bf k}$ dependence is in part explicit and in part implicit through  the dependence of  $ u_{{\bf k} m}$ on ${\bf k}$, which to linear order can be  written
\begin{equation}
u_{{\bf k}m}({\bf r}) = u^{(0)}_{m}({\bf r})+\sum_{\alpha} i k_{\alpha}  u^{(1)}_{\alpha,m}({\bf r})
\end{equation}
where the  $u\zero_m$'s are the $0^{th}$ order wavefunctions, namely the results of a $\Gamma$ only calculation, and the $u\one_{\alpha m}$'s the first order corrections. 
Since in the absence of a magnetic field, $u\zero_m$'s and $u\one_{\alpha m}$'s can be made real  the first order  correction to the density vanishes identically.

To the second order in ${\bf k}$,  $E({\bf k})$ becomes:
\begin{eqnarray}
E({\bf k}) & = & E_{KS}(\Gamma) + \frac{1}{2}{\bf k}^{2} +  \nonumber \\
   \sum_{\alpha \beta}k_{\alpha}k_{\beta}       & \Big[ &   \sum_{I,L}\Big( \langle u^{(0)}_{m}| r_{\alpha}| P_{I,L} \rangle \omega_{L} \langle P_{I,L} | r_{\beta}|u^{(0)}_{m} \rangle  \nonumber \\
        & - & \frac{1}{2}\langle u^{(0)}_{m}|  r_{\alpha}r_{\beta}| P_{I,L} \rangle \omega_{L} \langle P_{I,L} |u^{(0)}_{m}\rangle \nonumber \\
  &- &   \frac{1}{2}\langle u^{(0)}_{m}| P_{I,L} \rangle \omega_{L} \langle P_{I,L} |  r_{\alpha}r_{\beta}|u^{(0)}_{m} \rangle \Big)\Big] +  \nonumber \\
   \sum_{\alpha \beta}k_{\alpha}k_{\beta}  &\Big[& \langle i  u^{(1)}_{\alpha m}|-i \nabla_{\beta} |u^{(0)}_{m} \rangle + \langle u^{(0)}_{m}|-i \nabla_{\beta} | i u^{(1)}_{\alpha m} \rangle    \nonumber \\
         & +  & \sum_{I,L}\Big(\langle i u^{(1)}_{\alpha m}| -ir_{\beta}| P_{I,L} \rangle \omega_{L} \langle P_{I,L} |u^{(0)}_{m} \rangle \nonumber \\
         & +  & \langle i u^{(1)}_{\alpha m}| P_{I,L} \rangle \omega_{L} \langle P_{I,L} | -ir_{\beta}|u^{(0)}_{m} \rangle \nonumber \\
  & + & \langle u^{(0)}_{m}| -ir_{\beta}| P_{I,L} \rangle \omega_{L} \langle P_{I,L} |i u^{(1)}_{\alpha m} \rangle \nonumber \\
  & + & \langle u^{(0)}_{m}| P_{I,L} \rangle \omega_{L} \langle P_{I,L} |  -ir_{\beta}| i u^{(1)}_{\alpha m} \rangle \Big) \nonumber \\
                   & + & \langle u\one_{\alpha m}|H\zero_{KS} - \varepsilon\zero_{m}| u\one_{\beta m} \rangle\Big]
\label{eq: pert}
\end{eqnarray}
Once the $u\zero_{m}$'s are calculated, the $u\one_{\alpha m}$'s can be determined by minimizing the second order correction of the functional, $E\two$, relative to the $u\one_{\alpha m}$'s. The minimization procedure is to be supplemented by the restricted orthonormality condition \cite{Gonze}
\begin{equation}
\langle u\zero_{m} | u\one_{\alpha n} \rangle = 0 \qquad \forall \;  m,n  \; and \; \alpha
\end{equation} 

It can be seen that $E\two$ has the same structure as the standard form of the variational DFPT \cite{Putrino}. 
The only peculiarity is that  the quadratic terms  describing the effect of the change in $\rho$ due to the perturbation are absent.
Such a simplification is a consequence of the nature of the perturbation and occurs also, for instance, in the case  of the  calculation of NMR chemical shifts \cite{Putrino}.
In the DFPT the terms that are linear in $\{u\one_{\alpha m}\}$ express the perturbative coupling \cite{Putrino}.
In our case for a purely local PP the perturbation appears as a coupling to the current operator $-i{\bf \nabla}$, as is to be expected  from a ${\bf k}\cdot {\bf p}$ scheme.
Since it is well known that a non-local PP also affects the current operator, it is not  surprising that in the non-local case some extra perturbation terms appear.
We note that as written in Eq.(\ref{eq: pert}) the ${\bf k}$ expansion has only a formal character, since the position operator, in the non-local PP terms, is ill defined in a periodic system.
If however we express  the full non-local PP projectors in the reciprocal space and then expand in ${\bf k}$ the non-local  energy terms, we arrive at  well-defined expressions that can be straightforwardly calculated \cite{next}.
In practice we first perform a $\Gamma$ only calculation and then,  using the variational DFPT module developed in our group \cite{Putrino}, we perform three separate variational calculations to determine $\{u\one_{x m}, u\one_{y m}, u\one_{z m}\}$.
The calculations of $\{u\zero_m\}$ and $\{u\one_{\alpha m}\}$ can take full advantage of the fact that these quantities are real. Once the $\{u\one_{\alpha m}\}$
 are known, $\Delta_{\alpha \beta}=\delta^{2}E/(\delta k_{\alpha}\delta k_{\beta})|_{{\bf k}=0}$ can be evaluated and summing over the BZ we arrive at the following energy estimate 
\begin{eqnarray}
E &=& E_{\Gamma}+\frac{1}{2}\sum_{BZ}\sum_{\alpha\beta}k_{\alpha}k_{\beta}\Delta_{\alpha \beta}
\end{eqnarray}

Therefore, the computational effort required  by the whole procedure is  comparable to the optimization of four independent sets of  wavefunctions \cite{Putrino} and it does not depend on the number of $k$-points used in the BZ sampling.

We tested the efficiency of the method on the crystalline systems of Ar and Si. For these two systems we have considered supercells of increasing size and compared the $\Gamma$ only calculation with and without  the perturbation theory correction to the exact result obtained performing a full BZ sampling in the primitive cell.

\begin{figure}[t!]
\epsfxsize= 9.5 truecm
\centerline{\epsffile{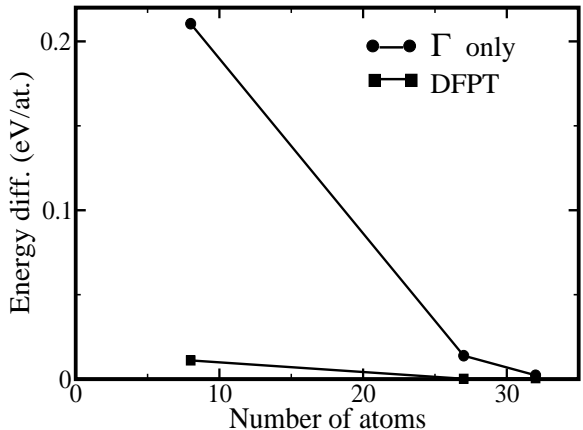}}
\renewcommand{\baselinestretch}{1.0}
\vspace{0.1truecm}
\caption{\em Total energy  vs the number of atoms in  Ar supercells. The energy differences are taken  with respect to the exact result corresponding to the full BZ sampling. Circles represent  differences due to $\Gamma$ only calculations, squares show the effects of DFPT correction for a BZ sampling generated by $(4 \times 4 \times 4)$ Monkhorst-Pack \cite{Monkhorst} mesh. In both calculations an  energy cutoff of 80 Ry was used. }
\label{fig: ARconv}
\end{figure}

The Ar $fcc$ unit cell, with 9.94 a.u. lattice constant, contains one atom and 8 valence electrons. We describe this system by means of  a Goedecker PP \cite{goedecker} and the LDA functional. 
The converged total  energy  is achieved  by the  standard CPMD \cite{cpmd} wavefunction optimization, over a $(8 \times 8 \times 8)$ Monkhorst-Pack (MP) \cite{Monkhorst}  mesh in the BZ of the primitive cell.
Calculating the total energy with the $\Gamma$ point only produces  errors which decreases with the system size, as shown in Fig. \ref{fig: ARconv}.
As expected in this case, since Ar bands are rather flat, the ${\bf k} \cdot {\bf p}$ approach converges very rapidly with the system size.

\begin{figure}[t!]
\epsfxsize= 10.3 truecm
\centerline{\epsffile{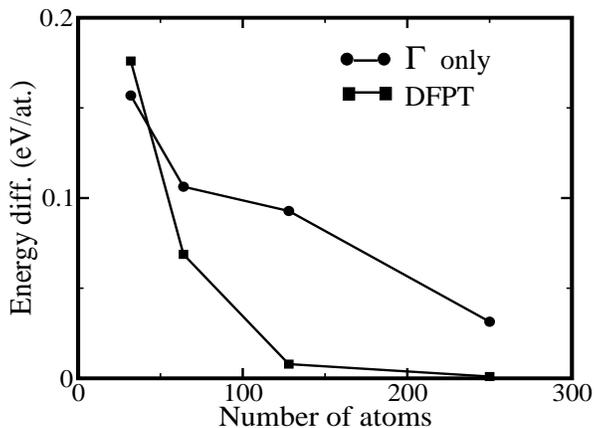}}
\caption{\em Total energy vs the number of atoms in Si supercells. The tech
nique details are the same as in Fig \ref{fig: ARconv}, but for the energy cutoff  which is 20 Ry.}
\label{fig: conv}
\end{figure}

We turn now to the study of the prototypical semiconductor, Si.
The Si unit cell contains 2 atoms and 8 valence electrons.
In this case we use the Troullier-Martins norm-conserving PP \cite{truiller} and  the Ceperley-Alder local density approximation to the exchange-correlation term \cite{Ceperley}.
 Convergence of total energy for this unit cell is achieved by an integration over the $(15\times 15\times 15 )$ MP  mesh.
As shown in Fig. \ref{fig: conv},  the DFPT correction produces an important improvement, such that the total energy error is reduced by one order of magnitude for 128-atom size, and to less than 0.001\% in the case of the 250-atom supercell. 
 Of course since the Si bands are much broader than those of Ar, convergence for Si is slower. 

\vspace{0.2 truecm}

Unfortunately the calculation of $\Delta_{\alpha \beta}$ by DFPT provides only the estimation of the total energy, which depends on the sum over the occupied state eigenvalues, and the contribution of the individual states cannot be separated. 
This is however necessary if one wishes to calculate the electronic density of states.
We show here that a very good electronic density of states  can be obtained if for any ${\bf k}$ one diagonalizes the ${\bf k}$ dependent KS Hamiltonian
\begin{eqnarray}
H({\bf k}) &=& H(0)
           +  \frac{1}{2}{\bf k}^{2} - i{\bf k}\cdot {\bf \nabla} \nonumber \\ 
          & + &  \sum_{I,L}e^{-i{\bf k}\cdot{\bf r}}|P_{IL}\rangle \omega_{L}\langle P_{IL}|e^{+i{\bf k}\cdot{\bf r}} \nonumber \\ 
           &-&  \sum_{I,L}|P_{IL}\rangle \omega_{L}\langle P_{IL}|
\end{eqnarray}
in the Hilbert space  spanned by the vectors $\{u\zero_{m} \}$ and $\{\tilde{u}\one_{{\bf k} m} \}$, which has dimension 2N, where
\begin{equation}
\tilde{u}\one_{{\bf k}m} = i k_{x} u\one_{x m}+ i k_{y} u\one_{y m}+ i k_{z} u\one_{z m}.
\end{equation}
These vectors are orthogonal to the subspace $\{ u\zero_{m} \}$ but they are not mutually orthogonal. 
We perform the orthonormalization  by means of the L\"owdin method, which requires  the overlap matrix
\begin{equation}
S_{m,n}({\bf k}) = \langle \tilde{u}\one_{{\bf k}m} | \tilde{u}\one_{{\bf k}n}\rangle = \sum_{\alpha \beta} k_{\alpha} k_{\beta} \langle u\one_{\alpha m}|u\one_{\beta n} \rangle
\end{equation} 
to be calculated for each ${\bf k}$ vector. The matrices $\langle u\one_{\alpha m}|u\one_{\beta n} \rangle$ can be calculated only once and then stored for later use.
Similarly it can  easily  be shown that most of the matrix elements of $H({\bf k})$ can be expressed in terms of ${\bf k}$ independent matrices, which  can be calculated outside the ${\bf k}$ loop.
The only exception is the non-local PP term, which requires the evaluation of the projectors $\langle P_{IL}|e^{i{\bf k}\cdot r}|u\zero_{m}\rangle$ and $\langle P_{IL}|e^{i{\bf k}\cdot r}|u\one_{\beta m}\rangle$.
However, fully in the spirit of our method, we can expand these projectors to second order in ${\bf k}$, which reduces the evaluation to a linear combination of ${\bf k}$ independent quantities.

\begin{figure}[h!]
\epsfxsize= 7.5 truecm
\centerline{\epsffile{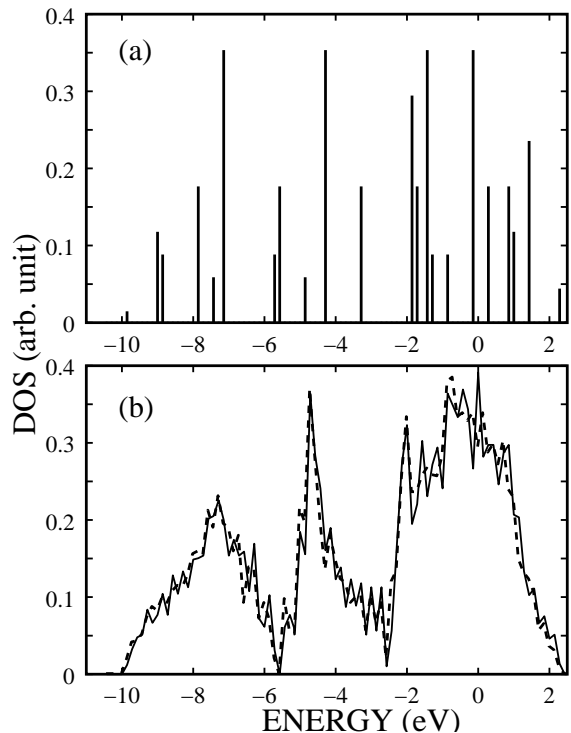}}
\caption{\em Electronic density of state calculated for the Si supercell containing 128 atoms. a) $\Gamma$ only DOS, b) Comparison between the standard CPMD (dashed) and the DFPT (solid) results, by using a $(4 \times 4 \times 4)$ MP mesh.}
\label{fig: dosmk4}
\end{figure}

We have tested our method on Ar and Si. Once again, for Ar the convergence is very fast and already for $\;$the  32-atom supercell, the ${\bf k} \cdot {\bf p}$ approach gives results that are practically indistinguishable from the real ones.
We illustrate here the more interesting case of Si. 
In Fig.$\;$\ref{fig: dosmk4}(a) we report the $\Gamma$ point only DOS, as resulted for a 128-atom Si supercell.   
\begin{figure}[t]
\epsfxsize= 7.5 truecm
\centerline{\epsffile{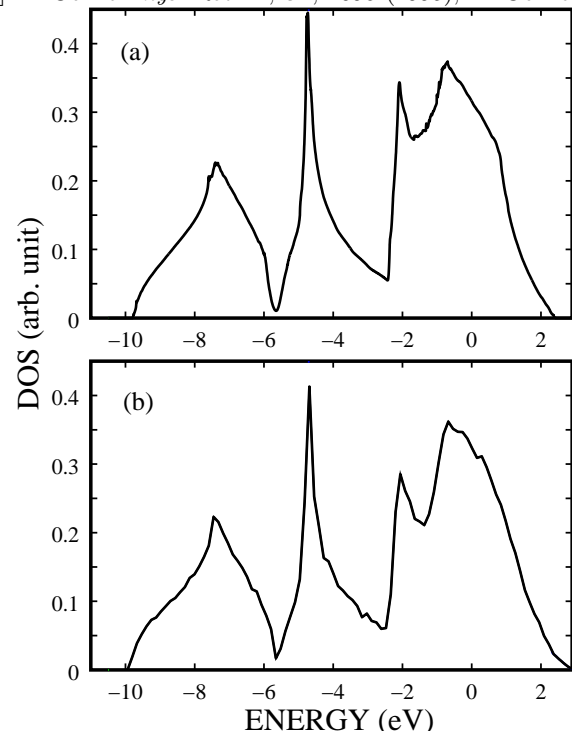}}
\caption{\em Si  electronic density of states: a) conventional fully converged calculation for a 2-atom unit cell with 8000 k-points, b) DFPT for a 128-atom supercell with 4096 k-points. In both cases the integration in the BZ by the tetrahedron method has been used.}
\label{fig: bestd}
\end{figure} 
It is easily seen that this is a caricature of the real DOS.
For the same system a much better description can be obtained  using a $(4 \times  4 \times 4)$ MP  mesh, as shown in Fig. \ref{fig: dosmk4}(b).
In the same picture we   demonstrate that our method gives  accurate results and that the use of the expansion in the evaluation of the non-local PP introduces  only small  errors.
While the result of the full calculation and of the ${\bf k} \cdot {\bf p}$ method  are comparable, the latter is computationally more efficient by at least two orders of magnitude. 
This allows us to calculate easily the electronic DOS for the 128-atom supercell including 4096 ${\bf k}$ points on an Intel Xeon PC.
The result is compared in Fig.\ref{fig: bestd} with the DOS obtained by a standard method  on the primitive cell of two atoms and with 8000 equally distributed ${\bf k}$ points. The only symmetry used is inversion.
For both calculations the tetrahedron method \cite{tetrahedron} was used in order to perform the ${\bf k}$ point integration in the BZ.
The  very good agreement between the two curves  demonstrates the usefulness of our approach.

In conclusion, we have shown that even for systems that are  by present computational standards  modest in size, the ${\bf k} \cdot {\bf p}$ method implemented here is accurate and computationally efficient.
It is therefore now possible to perform efficient {\it ab-initio} molecular dynamics calculations with fully converged  BZ sampling.

We are very grateful to the 
 Max Planck Instit\"ut fur Festk\"orperforschung,
  (Stuttgart, Germany) for its support during the initial phases of this research.



\end{document}